# On a method for the analysis of compulsive phase mixing and its application in cosmogony of galaxy superclusters

*S.N. Nuritdinov, A.A. Muminov*
*National University of Uzbekistan*

## ABSTRACT

In this paper, we study the strong non-stationary stochastic processes that take place in the phase space of self-gravitating systems at the initial non-stationary stage of their evolution. The numerical calculations of the compulsive phase mixing process were carried out based on the model of chaotic impacts, according to which the initially selected phase volume experiences random impacts of a different and complex nature.

## 1. INTRODUCTION

As known, at the initial non-stationary evolution stage of a collisionless self-gravitating system, phase mixing always takes place, which tends to bring the system closer to a quasistationary state in a regular field (Binney et al. 2008; Antonov et al. 1995). The statistical mechanics of stronger non-stationary case was first studied by Lynden-Bell (Lynden-Bell 1967), who called the corresponding physical process "violent relaxation". After the classical work of Lynden-Bell, a statistical study of "violent relaxation" became the main topic of discussion, where discussed the formation of star clusters, galaxies and their clusters. However, by numerical experiments (Madsen 1987; Merritt et al. 1989) it turned out that the predicted final coarse-grained distribution takes place only under extremely non-stationary initial conditions and only in one-dimensional analysis of this process. It should be noted that the work (Antonov et al. 1995) proposed a classification of phase mixing depending on the degree of initial non-stationarity $(2T/|U|)_0$ of the system, where T and U are the components of the kinetic and potential energy of the system, respectively. Within the framework of this work, we study the stochastic evolution of a selected volume element in four and six-dimensional phase spaces of a system with its most non-stationary state $(2T/|U|)_0 \ll 1$, which corresponds to compulsive phase mixing. Violent relaxation of Lynden-Bell is a special case of compulsive phase mixing. In compulsive phase mixing process, especially in the case of violent relaxation, the total gravitational field of the system changes stochastically at any point of the space over time, because of this, any small phase volume of the system will not be able to evolve independently. Thus, the phase volume is subjected to numberless random "impacts" (influences), which have different nature and intensity. We are interested in the statistical effect - the result of many influences. Calculations show that the initial random phase volume is statistically elongated on average in one direction, which confirms Lynden-Bell's predictions (Lynden-Bell 1967). Next, we are interested in this question in the case of a space with an odd dimension equal to 3. In this case, the statistical effect leads to the formation of a flat, almost disk-like formation. This result can be applied, in principle, to some astrophysical and cosmological processes. For instance, the presence of a turbulent stage in the evolution of the Universe (Gurevich, Chernin 1969; Peebles 1970) can probably statistically on average lead to the formation of galaxy protosuperclusters.

## 2. PHASE MODEL OF CHAOTIC IMPACTS

Let in the initial state there is an element of phase volume

$$\vec{q}_1^2 = 1, \qquad \vec{q} = \vec{q}(\vec{r}, \vec{v}) = 1. \tag{1}$$

If at time $t_1$ a random impact on a selected element of the phase volume is described by a random matrix $P_1$, then after the first impact we have $P_1 \vec{q} = \vec{q}_1$. From here we find $\vec{q} = P_1^{-1} \vec{q}_1$ and substitute in (1). We get

$$\vec{q}_1^T \left( P_1 P_1^T \right)^{-1} \vec{q}_1 = 1 \tag{2}$$

where the symbol T means transposition of the matrix. Expression (2) is the equation of an ellipsoid whose semi-axes are determined by the eigenvalues of a random matrix $P_1 P_1^T$. Thus, after the next impact in (2) we add the effect of the second «push»

$$\vec{q}_2^T \left( P_2 P_1 P_1^T P_2^T \right)^{-1} \vec{q}_2 = 1 \tag{3}$$

where $\vec{q}_2 = P_2 \vec{q}_1$ and $\vec{q}_1 = P_2^{-1} \vec{q}_2$. Therefore, the degree of extension of the phase element at time $t_n$ is characterized by the eigenvalues of the matrix $K_n$

$$K_n = P_n P_{n-1} ... P_2 P_1 P_1^T P_2^T ... P_{n-1}^T P_n^T. \tag{4}$$

Thus, the speed of stretching or compression of the initially selected phase volume in time is determined by the following relationship:

$$\eta_{ni} = \frac{1}{n} \ln \left( \frac{\lambda_{ni}}{\lambda_{ni+1}} \right) \tag{5}$$

where $\lambda_{ni}$ (i = 1,2…) is eigenvalues of matrix $K_n$. To determine the statistical characteristics of the process, we need to repeat the calculation of $\eta_{ni}$ many times and average over the number of realizations m, therefore, we can calculate $\eta_{ni}$ and its dispersion $\sigma_n (\eta, m)$

$$\sigma_n (\eta, m) = \left[ m^{-1} \sum_{i=1}^{m} \eta_{ni}^2 - \left( m^{-1} \sum_{i=1}^{m} \eta_{ni} \right)^2 \right]^{1/2} \tag{6}$$

It should be noted that in the process of studying the behavior of the test phase volume, various matrices can be used that satisfy the following conditions and are randomly selected depending on the parameter p, which varies in the range (0,1):

$$A\vec{q} \neq B\vec{q} \neq \vec{q}, \qquad AB \neq BA, \qquad |\det A| = |\det B| = 1. \tag{7}$$

Here, the meaning of first two expressions is clear and the last requirement follows from the Louisville theorem, which ensures the conservation of the phase volume in time. All matrices used in the calculations of the evolution of the selected phase volume satisfy condition (7)

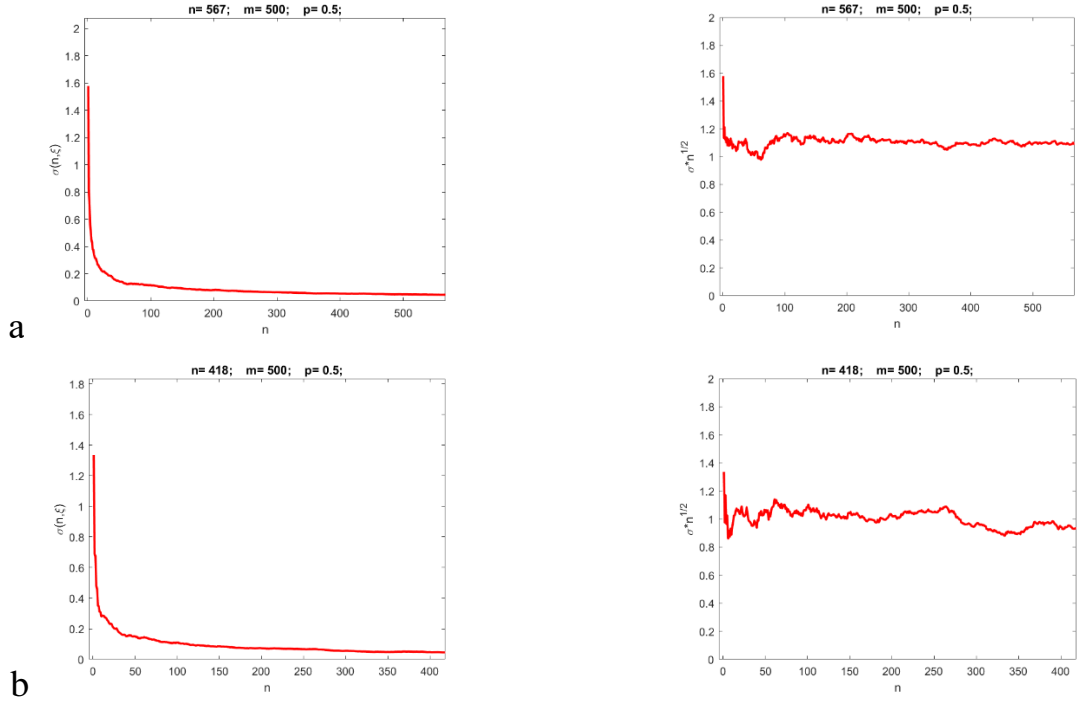

$$A(k_i) = \begin{pmatrix} k_1 & 0 & 0 & 0 \\ 0 & k_2 & 0 & 0 \\ 0 & 0 & k_1^{-1} & 0 \\ 0 & 0 & 0 & k_2^{-1} \end{pmatrix}, \quad B(h_i) = \begin{pmatrix} 2h_1 & 0 & 0 & h_1 \\ 0 & h_1^{-1} & 0 & 0 \\ 0 & 0 & h_2 & 0 \\ h_2^{-1} & 0 & 0 & h_2^{-1} \end{pmatrix}, \quad C(\alpha) = \begin{pmatrix} \cos\alpha & 0 & 0 & \sin\alpha \\ 0 & 1 & 0 & 0 \\ 0 & 0 & 1 & 0 \\ -\sin\alpha & 0 & 0 & \cos\alpha \end{pmatrix}.$$

*Fig. 1. Dependences of $\sigma$ and $\sigma \cdot \sqrt{n}$ on step $n$*

In order to calculate the statistical stretching of the selected phase volume, a corresponding computer code was written, with the help of which the values of the semi-axes of the four-axis ellipsoid and the dispersion of the elongation of the test phase volume were determined. Since we are dealing with a large number of random processes, we were interested in checking the central limit theorem of probability theory (Tutubalin+1992).

In calculating the evolution of the test phase volume, we selected two modes: in the first, two random matrices A and B were taken as the acting matrices. If the value of the parameter $p \leq 0.5$ then $P_i = A$ and $P_i = B$ in the opposite cases; in the second mode, three random matrices A, B, C were used, where $P_i = A$ when $p \leq 0.33$, $P_i = B$ with probability $p \leq 0.66$, in cases $p \leq 0.99$, $P_i = C$ is selected as the acting matrix.

According to our estimations, the number of realization in all calculations was taken to be m = 500, because more than this value, its contribution to the accuracy of calculations is almost not noticeable. Thus, after each «push», the eigenvalue of the matrix is calculated and the stretching velocity of the test phase volume is determined from them, then its elongation dispersion is calculated until its value is less than ε, where $\varepsilon = |\sigma_i(\eta,n) - \sigma_{i-1}(\eta,n)| = 10^{-6}$.

## 3. APPLICATION OF THE METHOD TO A THREE-DIMENSIONAL CASE

It is of some interest to consider a 3-dimensional case in ordinary physical space. The fact is, in the evolution of the Universe there is a period of gravitational turbulence (Weizsäcker 1951; Gamow 1952; Ozernoi & Chernin 1969), which is associated with dark matter and dark energy. On the other hand, in the 70s, Ya.B. Zel'dovich, considering the evolution of an element of the medium around which the velocities are arbitrary, believes that this element gradually takes a disk-like shape.

Ya. B. Zeldovich (Zeldovich 1970) connects these structures of the Universe with initially small adiabatic perturbations of density, velocity, and other parameters in a homogeneous Universe, and in order to explain the modern structure, the density perturbation amplitude in the epoch of recombination was about 0.05 medium density. His calculations showed that, due to nonlinear gas-dynamic effects, perturbations increase over time, and large regions of high density appear in the form of thin, flat layers of matter by the period of the formation of protogalaxies. Later, they were called pancakes (zeldovich+1978). In the central part of the pancake, the substance is highly compressed, and a shock wave arises in the cold gas flowing into the center. The temperature immediately after the front of the shock wave is extremely high, but then due to radiant cooling it should fall, and especially quickly in the central, most dense region. Here, already under the influence of gravitational and thermal instability, decay, fragmentation of the pancake into protogalactic gas clouds should occur, and it should turn into a cluster of protogalaxies. The identification with the cluster is not accidental: the characteristic mass of pancakes is $10^{13}$-$10^{14}$ M$_\odot$, that is, it coincides with the typical mass of galaxy clusters. This most important quantitative result of the theory is obtained from an analysis of the behavior of small perturbations in a hot Universe.

Based on the foregoing, it can be assumed that random forces and disturbances act on the selected element of the turbulent medium. Then it is necessary to determine the statistical effect of these random effects on a given element. Does this effect lead to the formation of just a flat formation or again to a state elongated along one axis?

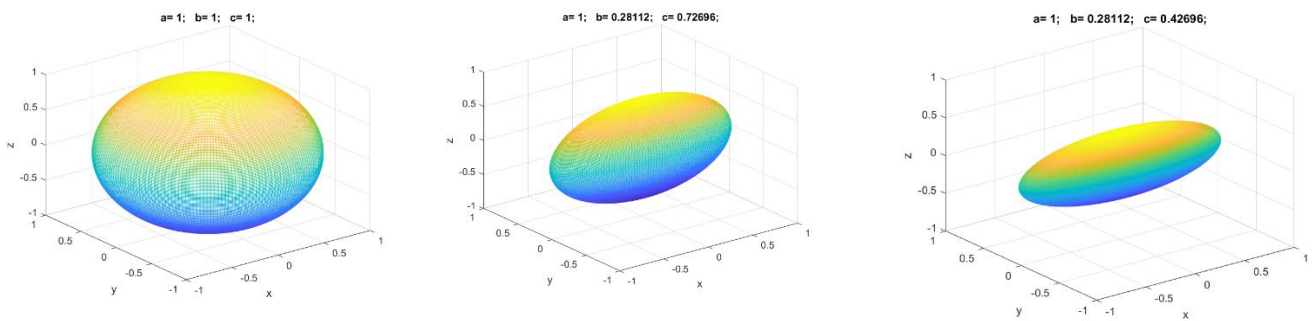

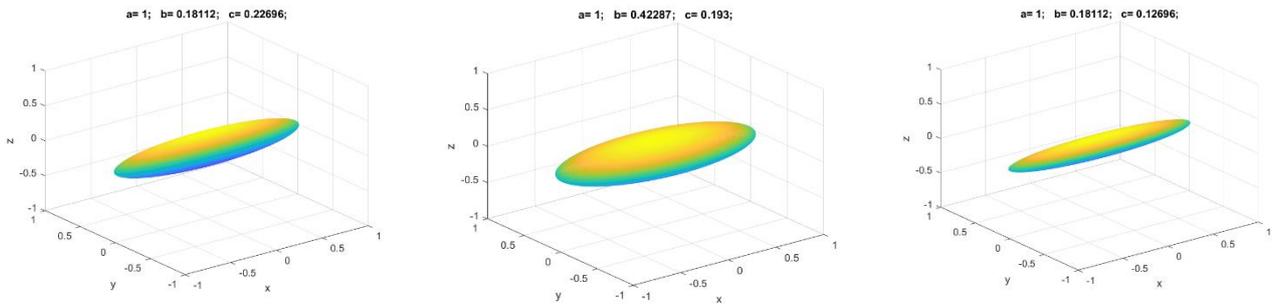

*Fig.2.* Behavior of sphere after a series of impacts.

Based on our model of random effects by random matrices, we investigated the behavior of the original element in time. As it turned out, after a series of random initial volume element gradually takes a disk-like shape (Fig. 2). So there is the formation of Zeldovich pancakes in the early stages of the formation of the universe. Note that in the numerical calculations we used the following random matrices:

$$A(k_i) = \begin{pmatrix} k_1 & 0 & 0 \\ 0 & k_2 & 0 \\ 0 & 0 & 1/k_1 k_2 \end{pmatrix}, \quad B(\alpha) = \begin{pmatrix} \cos\alpha & 0 & \sin\alpha \\ 0 & 1 & 0 \\ -\sin\alpha & 0 & \cos\alpha \end{pmatrix}.$$

These random matrices also satisfy condition (7).

## 4. CONCLUSION

In the framework of this work, the following results were obtained:

- a technique is proposed for analyzing the statistical effect of random effects on a selected arbitrary volume element in the case of four-dimensional phase space;

- with an increase in the number of random matrices, the characteristic time decreases noticeably;

- it is determined that the final form in the four-dimensional phase space is a filamentous structure elongated in one direction;

- the application of the method for studying random effects on a volume element in the case of three-dimensional space is carried out. Received a flat formation resembling "pancakes" Zeldovich;

- verified the fulfillment of the central limit theorem of probability theory, which obey the law $\sigma_i(\eta,n) \sim n^{-1/2}$.